\documentclass{iau}

\usepackage{amsmath}
\usepackage{graphicx}
\usepackage{multirow}
\usepackage{array}
\usepackage{multicol}
\usepackage{adjmulticol}
\usepackage{wrapfig}

\newcommand{\GG}{\mbox{$G$}}
\newcommand{\GBP}{\mbox{$G_{\rm BP}$}}
\newcommand{\GRP}{\mbox{$G_{\rm RP}$}}
\newcommand{\GRVS}{\mbox{$G_{\rm RVS}$}}
\newcommand{\GBPmGRP}{\mbox{$G_{\rm BP}-G_{\rm RP}$}}

\newcommand{\mci}[1]{\multicolumn{1}{c}{#1}}

\newcommand{\Teff}{\mbox{$T_{\rm eff}$}}

\newcommand{\EBV}{\mbox{$E(4405-5495)$}}
\newcommand{\RV}{\mbox{$R_{5495}$}}

\begin{document}

\lefttitle{J. Ma\'{\i}z Apell\'aniz}
\righttitle{\textit{Gaia} and massive stars in 2025}

\jnlPage{1}{7}
\jnlDoiYr{2025}
\doival{10.1017/xxxxx}

\aopheadtitle{Proceedings IAU Symposium}
\volno{402}
\editors{A. Wofford,  N. St-Louis, M. García \&  S. Simón-Díaz, eds.}

\title{\textit{Gaia} and massive stars in 2025}

\author{J. Ma\'{\i}z Apell\'aniz}
\affiliation{Centro de Astrobiolog\'{\i}a, Campus ESAC, CSIC-INTA. 28\,692 Villanueva de la Ca\~nada, Madrid, Spain}

\begin{abstract}
I present some of the highlights of the \textit{Gaia} mission on massive stars and discuss what the fourth data release (DR4) will bring in late 2026. In the first part of the contribution I describe the different types of \textit{Gaia} products available now and for DR4 and their caveats. In the second part I present the most significant results on massive stars regarding parameter determination, binaries, photometric variability, stellar groups, and the sample and spatial distribution.  
\end{abstract}

\begin{keywords}
Astrometry, Gaia, Galactic structure, Multiple stars, Photometry, Spectroscopy, Stellar clusters and OB associations,
Stellar variability
\end{keywords}

\maketitle

\section{The \textit{Gaia} mission: products and caveats}

$\,\!$ The \textit{Gaia} mission \citep{Prusetal16} took data for ten and a half years from 24 July 2014 to 15 January 2025, yielding the most complete ever high-dynamic-range, astrometric+photometric+spectroscopic survey of the whole sky. The third data release (DR3) took place in two phases in 2020 and 2022 \citep{Browetal21,Valletal23} and the next and fourth one (DR4) is expected to take place in December 2026\footnote{\url{https://www.cosmos.esa.int/web/gaia/release}}. Table~\ref{Gaiaproducts} summarizes the different data products available in DR3 and DR4. In this section we describe the characteristics and caveats of the different types. 

\subsection{Astrometry}

$\,\!$ \textit{Gaia}~DR3 astrometry is described in \citet{Lindetal21a,Lindetal21b} as part of the ``official'' early-DR3 (EDR3) papers from DPAC (\textit{Gaia} Data Processing and Analysis Consortium\footnote{\url{https://www.cosmos.esa.int/web/gaia/dpac/consortium}}). Further contributions from \citet{CanGBran21,Maizetal21c,Maiz22} have expanded and refined that analysis. The most important points that need to be considered for DR3 astrometry are:

\begin{itemize}
 \item Distances depend on the chosen prior \citep{Maiz05c,Lurietal18}. The commonly used \citet{Bailetal21} distances assume an underlying spatial distribution for low- and intermediate-mass stars. \citet{Pantetal25b} instead assumes a spatial distribution for Galactic OB stars \citep{Maiz01a,Maizetal08a}.
 \item Catalog parallaxes require a zero point correction \citep{Lindetal21b,Maiz22}. \citet{CanGBran21} derive a similar correction for proper motions.
 \item Catalog (or internal) parallax uncertainties are underestimated by up to a factor of three (\citealt{Fabretal21a,Maiz22}, see also \url{https://www.cosmos.esa.int/documents/29201/1770596/Lindegren_GaiaDR2_Astrometry_extended.pdf}).
 \item There is a residual systematic uncertainty ($\sigma_{\rm s} = 10.3\;\mu$as) that yields a minimum uncertainty for distances of $\approx d\%$, where $d$ is the distance in kpc \citep{Maizetal21c}.
 \item The residual systematic uncertainty has an angular correlation \citep{Lindetal18a,Maizetal21c}.
 \item Poor-quality data (RUWE $>$ 1.4 or six-parameter solutions) can still be used with a larger uncertainty \citep{Maiz22}.
\end{itemize}

\begin{table}[]
 \centerline{
 \begin{tabular}{lm{3.6cm}m{3.9cm}m{3.5cm}}
  \midrule
\mci{Type}        & \mci{Details}           & \mci{DR3}             & \mci{DR4}           \\
  \midrule
All               & Release dates+data span & 2020/2022, 34 months  & Dec 2026, 66 months \\ 
Astrometry        & Coordinates + parallaxes + proper motions       & 
                    Mean + uncertainties                            &
                    Mean + uncertainties + epoch data               \\  
Photometry        & \GG + \GBP + \GRP + \GRVS                       &  
                    Mean + uncertainties + epoch data ($\sim 1\%)$  & 
                    Mean + uncertainties + epoch data               \\  
Spectrophotometry & 3300-10\,200 \AA\ low + variable resolution     & 
                    Mean basis coefficients for $\sim 10\%$         &  
                    Mean + epoch basis coefficients for X$\%$       \\  
Spectroscopy      & 8450-8740 \AA\ $R\sim 11\,500$                  &  
                    Mean spectra for $\sim0.1\%$ + 
                    Mean RV for $\sim10\%$                          &  
                    Mean + epoch spectra + RV for X$\%$             \\  
  \midrule
 \end{tabular}
 }
 \caption{Summary of \textit{Gaia} data products for DR3 and DR4}
 \label{Gaiaproducts}
\end{table}

\vspace{-5mm}
\subsection{Photometry}

$\,\!$ \textit{Gaia}~DR3 photometry is described in \citet{Rieletal21} as part of the DPAC papers. In \citet{MaizWeil25} we reanalyse it and provide a model to use it when comparing to synthetic photometry:

\begin{itemize}
 \item \textit{Gaia} optical photometry is unparalleled in coverage, dynamic range and uniformity but the passbands have evolved with each data release (from DR1 to DR3 and likely DR4, see \citealt{Maiz17a,MaizWeil18}) likely as a result of water deposition in optical surfaces at the beginning of the mission \citep{Prusetal16} and differences in processing between data releases.
 \item \GG\ catalog magnitudes require a correction.
 \item \GG\ magnitudes in DR2 and DR3 require two different passbands with a cutoff at $\GG = 13.0$ and \GBP\ and \GRP\ magnitudes require two different passbands in DR2 with a cutoff at $\GG = 10.87$. This is likely a result of the electronic gate-based data processing \citep{Prusetal16}.
 \item If the \citet{MaizWeil25} procedure is followed it is possible to attain an absolute calibration between 3.4~and~5.7~mmag for \textit{Gaia}~DR3 photometry.
 \end{itemize}

\vspace{-5mm}
\subsection{Spectrophotometry}

$\,\!$ \textit{Gaia} spectrophotometry results from combining BP and RP prism data to cover the 3300-10\,200~\AA\ range. The combination of low and variable resolution complicates the interpretation of the results. The output is given as a series of coefficients of an expansion in basis functions, but the choice of basis functions partially determines the derived $f_\lambda(\lambda)$ and possibly introduces artifacts if the real $f_\lambda(\lambda)$ cannot be realistically expressed as a combination of the basis functions \citep{Weiletal20}. This is especially true for extinguished OB stars. Under those conditions, synthetic photometry and line measurements are not straightforward \citep{Weiletal23}. The reader is encouraged to read those papers before using the data. 

\vspace{-5mm}
\subsection{Spectroscopy}

$\,\!$ Data from the \textit{Gaia} Radial Velocity Spectrometer (RVS) has a resolution of 11\,500 and covers the Calcium triplet window, where few lines exist for OB stars other than the broad Paschen lines. Furthermore, very few spectra of massive stars were published in DR3 and the papers on ``hot stars'' in DR3 refers to mostly intermediate- and low-mass stars \citep{Blometal23,Fremetal23}. Hopefully in DR4 there will more RVS data for massive stars.

\vspace{-5mm}
\subsection{Multi-origin processed data products}

$\,\!$ \textit{Gaia} also provides a number of processed products obtained by combining different data types. An example is the \textit{Gaia} astrophysical parameters inference system (Apsis, \citealt{Bailetal13}), for which the results for OB stars in DR3 are described below.

\begin{figure}[t]
  \centerline{
  \hspace{-5mm}
  \includegraphics[width=0.53\textwidth]{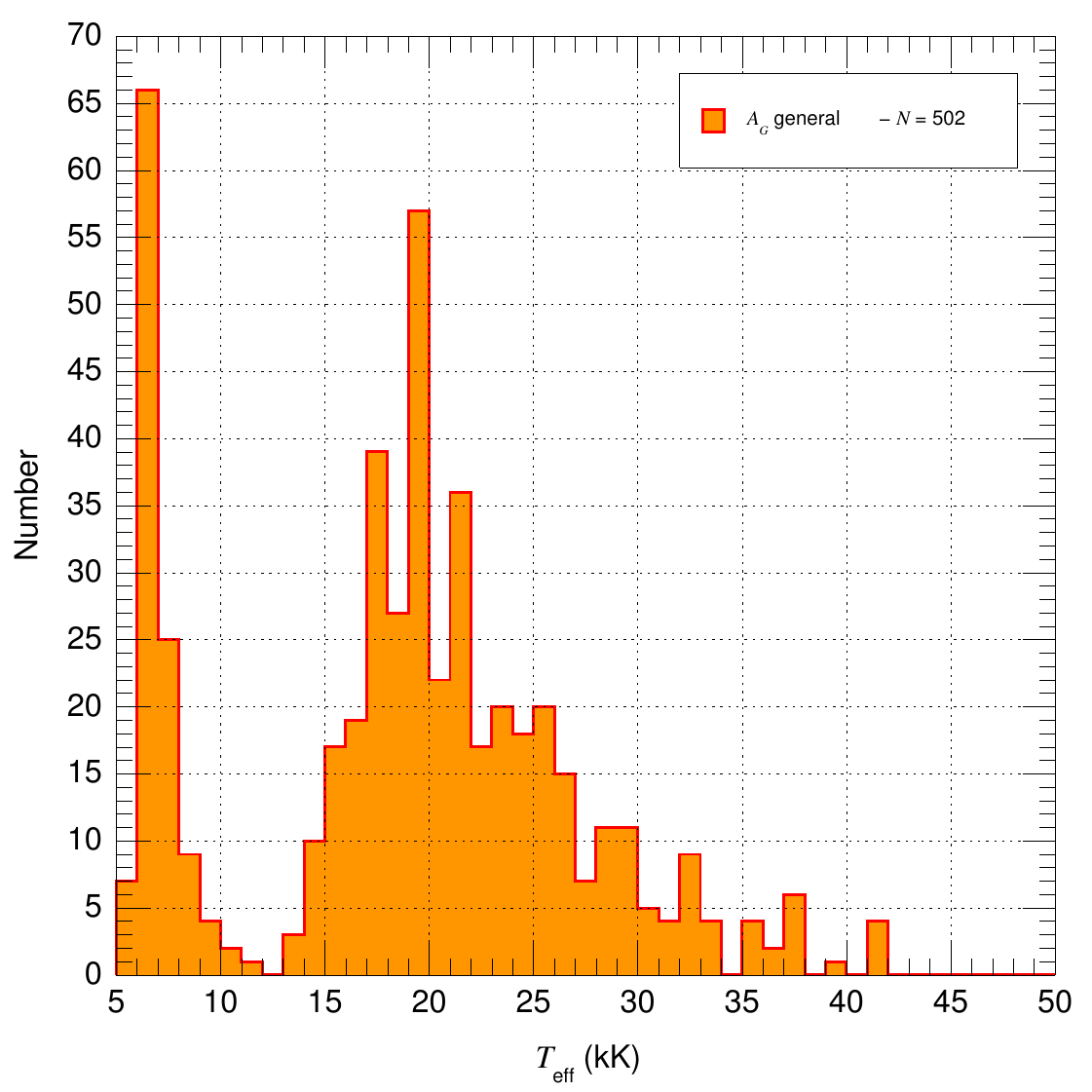} \
  \includegraphics[width=0.53\textwidth]{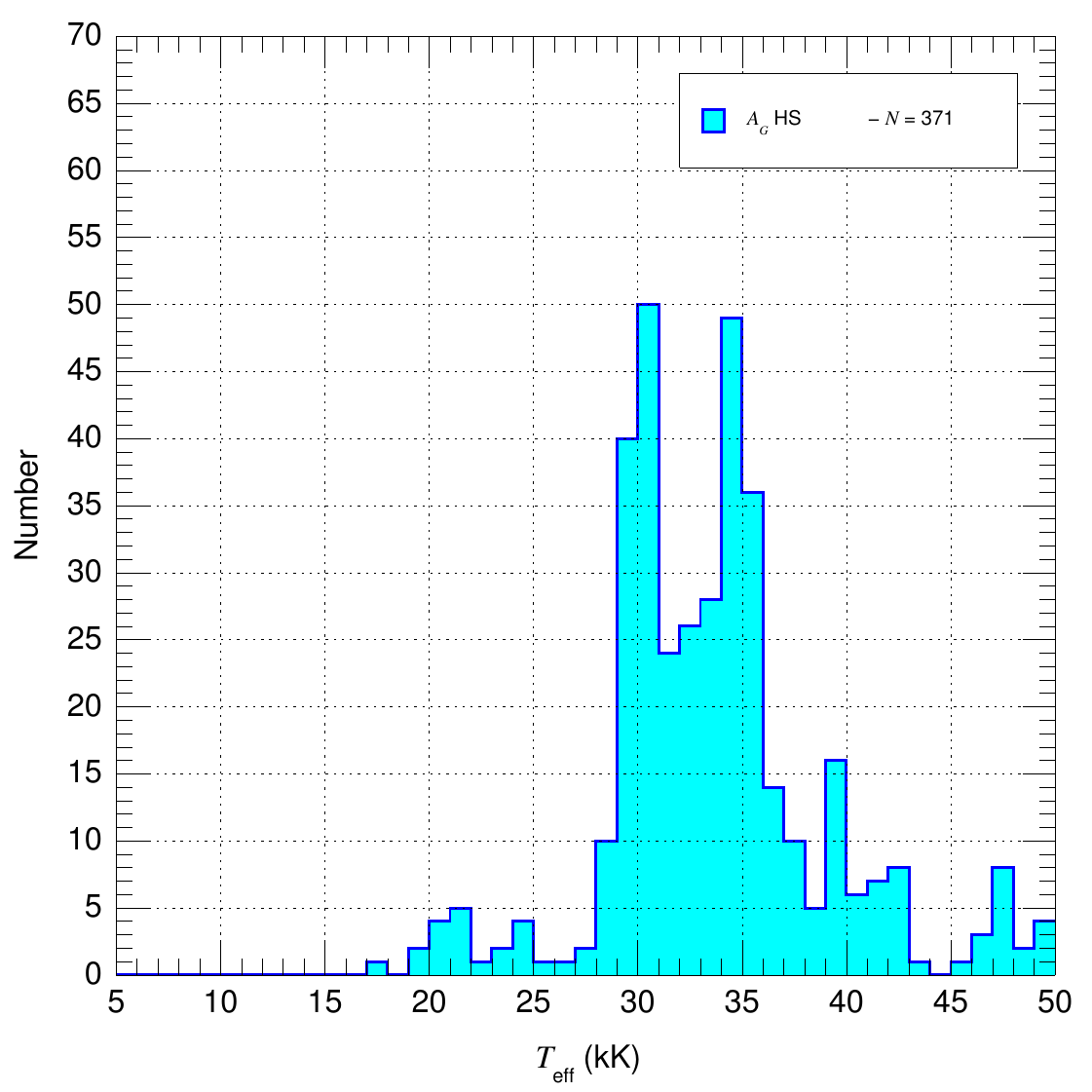} 
  }
  \caption{Values for \Teff\ determined by Apsis for 947 O stars with accurate GOSSS spectral classifications \citep{Maizetal11} using GSP-Phot (left) and ESP-HS (right).}
  \label{Apsis}
\end{figure}

\vspace{-5mm}
\section{Parameter determination}

$\,\!$ The DR3 DPAC determination of stellar parameters using Apsis is described in \citet{Creeetal23a,Foueetal23} and the specific case of ``hot stars'' (defined as those with $\Teff > 7.5$~kK) is discussed in \citet{Frem24}. There are different pipelines involved in Apsis, including a general one called GSP-Phot and another one for hot stars called ESP-HS. \citet{Frem24} shows poor results in the confusion matrix that compares the spectral types predicted by ESP-HS and the ones from Simbad (which have their own problems, see \citealt{Maizetal16}), with a large fraction of misclassified O and B stars, including a majority of O stars. I did a similar study (compatible with the \citealt{Frem24} results), partially presented as \citet{Maiz24} comparing the DR3 Apsis \Teff\ values with O stars with accurate GOSSS spectral types \citep{Maizetal11}. Figure~\ref{Apsis} shows how GSP-Phot only correctly identifies a few O stars as such, with a significant fraction being assigned $\Teff < 10$~kK. ESP-HS does a much better job but still underestimates \Teff\ for most O stars.

Therefore, at least for DR3, Apsis does a poor job at determining the properties for OB stars but why? There are at least two reasons:

[1] Apsis (and \textit{Gaia} in general) was designed mostly with cool stars in mind. To determine the properties of hot stars (a) the blue-violet range is much better that the Calcium triplet window of RVS and (b) the Balmer jump is difficult to measure with BP spectrophotometry because it is located close to its wavelength edge and also likely because the predetermined basis functions do not work well with extinguished OB stars.

[2] Apsis uses the \citet{Fitz99} family of extinction laws instead of other existing better choices and, more importantly, fixes the value of $R$ to 3.1, which only works properly for a minority of OB stars \citep{Maizetal14a,MaizBarb18}. As a consequence, the determined extinguished SED is inconsistent with the same \textit{Gaia} spectrophotometry used to establish it (see Fig.~9 in \citealt{Maiz24}). The incorrectly determined extinction quite likely interferes with an accurate \Teff\ determination.

\begin{figure}[t]
  \centerline{
  \hspace{-5mm}
  \includegraphics[width=0.53\textwidth]{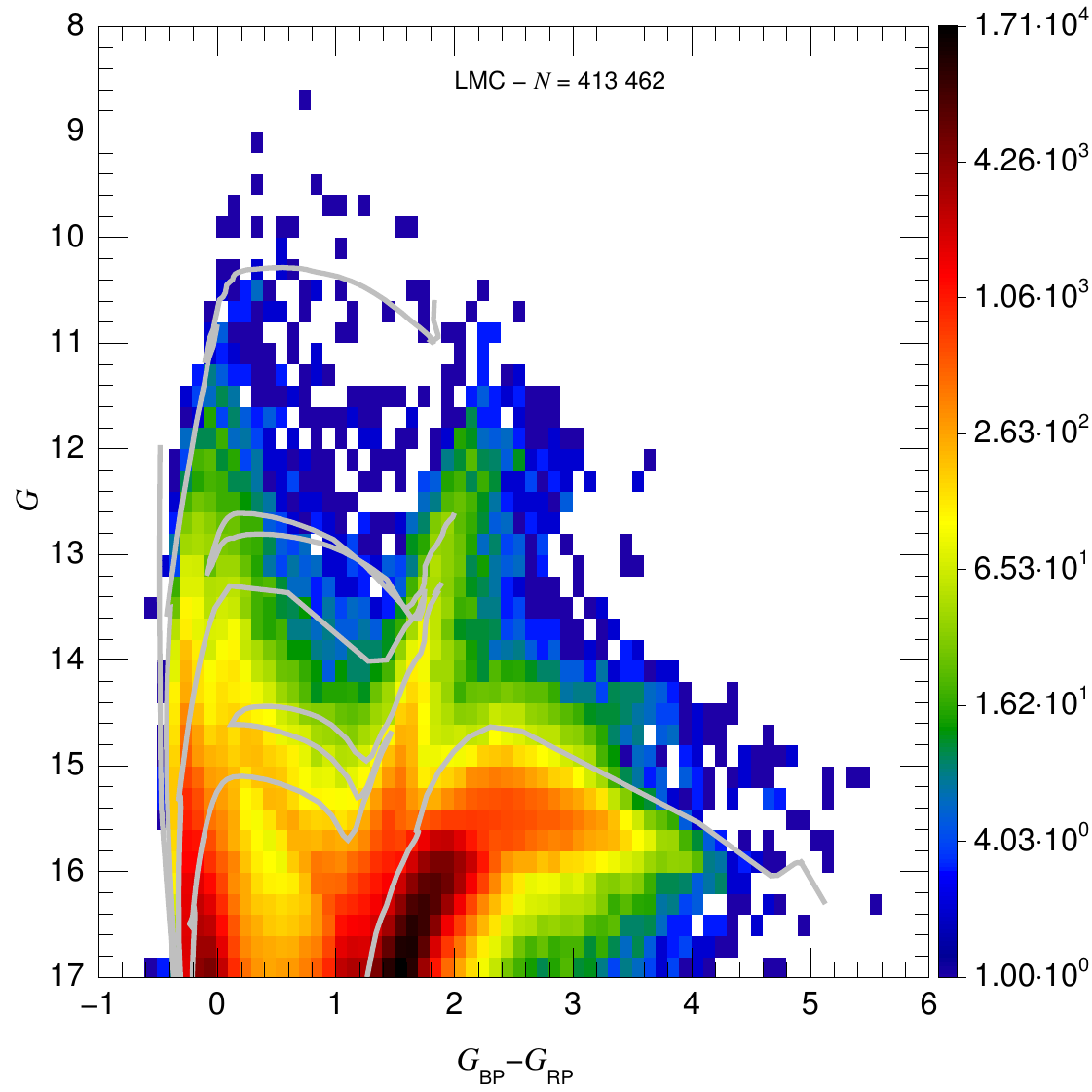} \
  \includegraphics[width=0.53\textwidth]{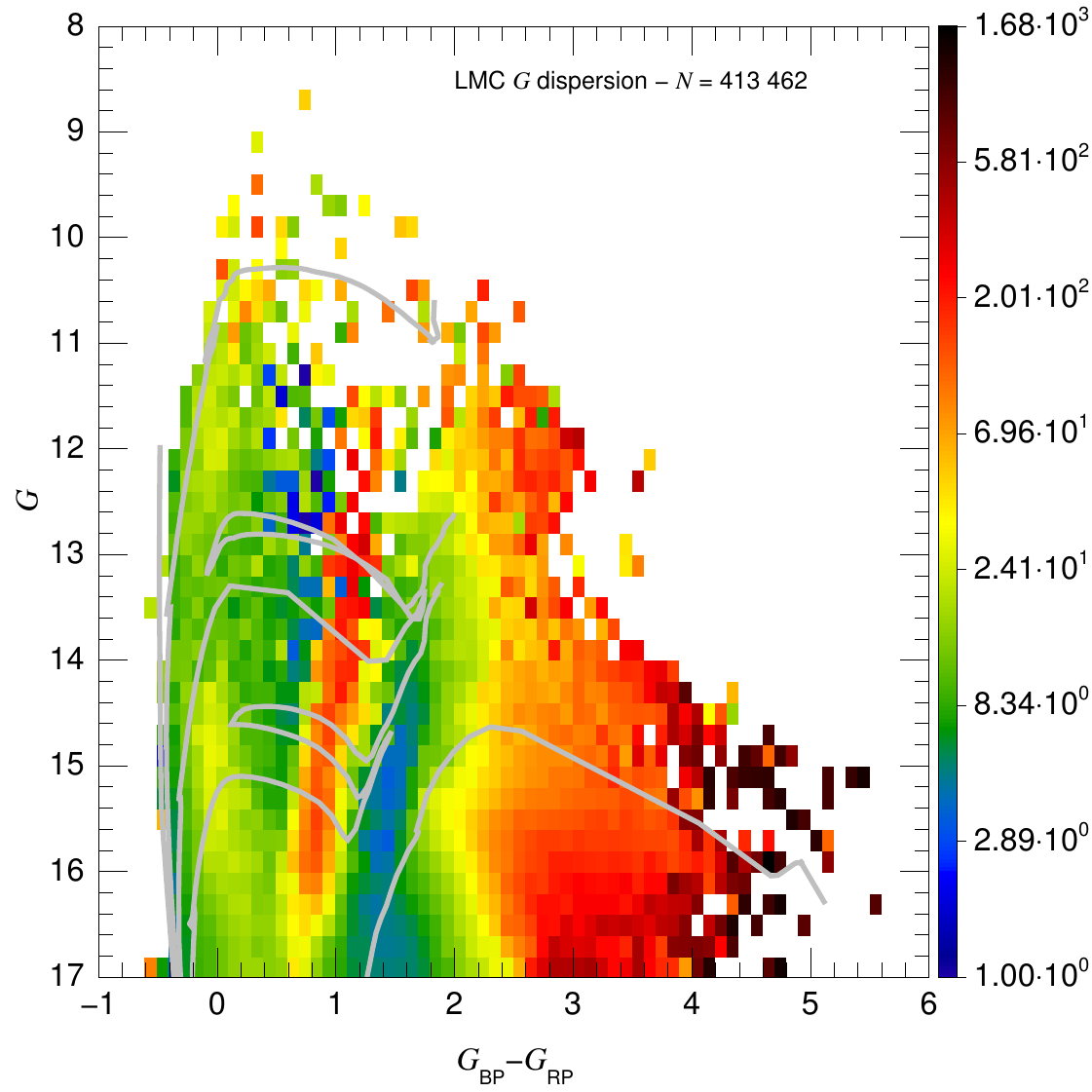} 
  }
  \caption{(left) CMD and (right) mean \GG-band photometric dispersion in mmag for the LMC \textit{Gaia}~DR3 sample of \citet{Maizetal23}. The most prominent variability regions correspond (in increasing \GBPmGRP) to Be stars, cepheids, and long-period variables, respectively.}
  \label{variability}
\end{figure}

\section{Binaries}

$\,\!$ The largest DR3-based study of binaries is \citet{Mowletal23}, who analysed over 2 million eclipsing binaries from the \citet{Rimoetal23} sample, see below. \textit{Gaia}~DR3 typically included 50 photometric epochs (the exact number depends primarily on sky position) and the high photometric accuracy, spatial resolution, and stability of \textit{Gaia} photometry is a good complement to other surveys such as TESS, which has a much better sampling but poorer spatial resolution and long-term photometric accuracy. See \citet{Holgetal25a} for an example of how to combine the best of both worlds to study massive eclipsing binaries. DR4 will provide $\sim 100$ photometric epochs for a much larger sample and should allow the discovery and study of even more eclipsing binaries (massive and non-massive).

Of course, the most interesting massive binaries are those that are expected to be discovered through the combination of epoch astrometry and radial velocities, especially those with a BH or NS component where the motion of the photocentre is most prominent \citep{Andretal19,Jansetal22}. However, as \textit{Gaia}~DR3 published very limited epoch data, only two systems with BHs were discovered there \citep{ElBaetal23b,ElBaetal23a}. A third one was discovered using pre-release DR4 astrometry \citep{Panuetal24}. DR4 is expected to yield a large number of systems with dark companions and a review similar to this one in the next beach meeting will likely be very different.

\section{Photometric variability} 

$\,\!$ \textit{Gaia}~DR3 contained epoch photometry for only $\sim 1\%$ of the sample, with the targets selected on the basis of their (suspected) variability. That epoch photometry is the basis for the classification of 12.4 million photometric variables by \citet{Rimoetal23}, which included classes (mostly) specific to massive stars such as $\alpha$~Cyg, $\beta$~Cep, and Be (+ other emission line stars) and other more generic ones such as eclipsing binaries (see above) and long-period variables that included some massive stars. A comparison with TESS data reveals that \textit{Gaia} photometry is highly useful for the classification of variable stars \citep{HeyAert24}.

One problem with using the \citet{Rimoetal23} results is that their sample is heavily biased by design towards highly variable stars, leaving no variability data for most stars and without the possibility of doing global mean-variability studies for most regions of the CMD, at least until DR4 becomes available. To address those issues, \citet{Maizetal23} devised a method to obtain the photometric dispersion (or variabilty amplitude) in the three \GG+\GBP+\GRP\ bands for the sample of \textit{Gaia} stars with $\GG > 17$~mag and five-parameter astrometric solutions (the majority of them). An example of their results is shown in Fig.~\ref{variability}, where the mean \GG\ photometric dispersion for 413\,462 bright LMC stars is shown, allowing for the identification of variable and non-variable stars alike. As a comparison, less than 13\% of those stars are in \citet{Rimoetal23}.

\section{Stellar groups}

$\,\!$ \textit{Gaia} has had a tremendous impact on the study of stellar groups (clusters and associations). One important area has been the census of open clusters and the determination of their characteristics (\citealt{CanGetal18,CasGetal20,CasGetal22,HuntReff23}, to name some of the relevant studies). Another one has been the determination of the internal kinematics of stellar groups, with the evidence piling up against a monolithic model of star formation of clusters followed by a posterior dispersion and in favour of a hierarchical model with stars being born both in clusters of different sizes and in looser associations \citep{Carr20,Wrig20,Wardetal20,QuinWrig22,Vaheetal23}.

\begin{figure}[t]
  \centerline{
  \hspace{-5mm}
  \includegraphics[width=0.53\textwidth]{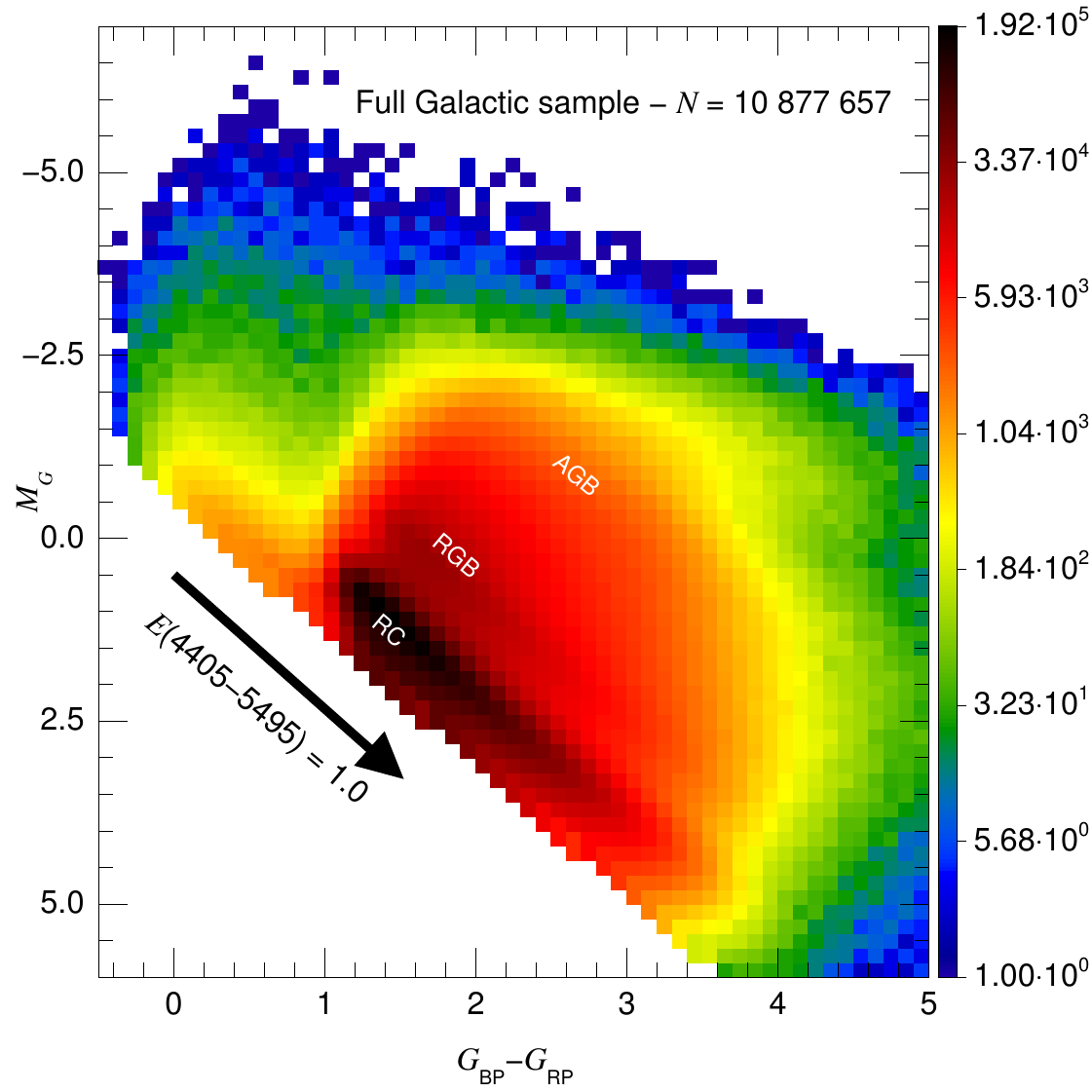} \
  \includegraphics[width=0.53\textwidth]{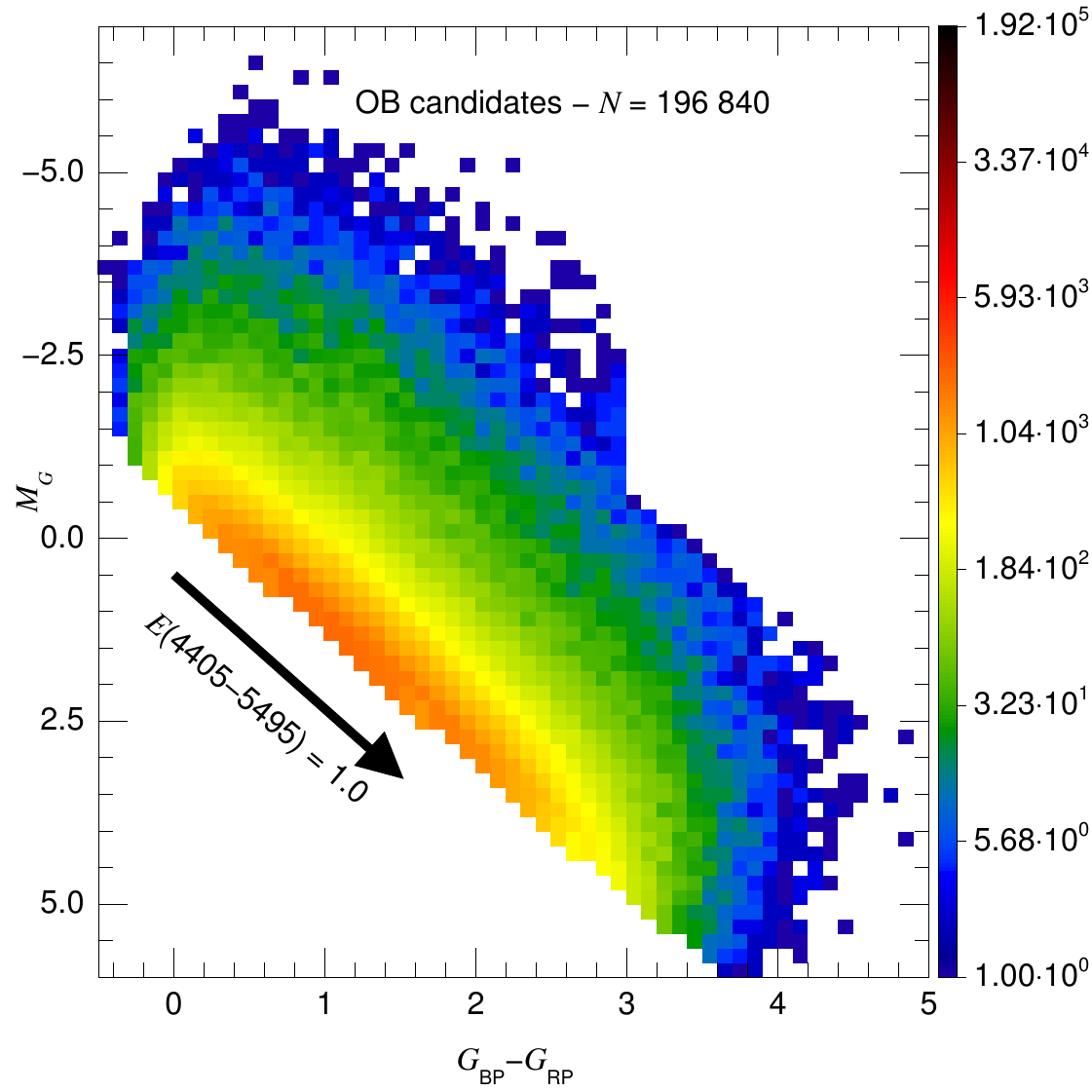} 
  }
  \caption{(left) CAMD for Galactic stars with good-quality \textit{Gaia}~DR3 + 2MASS data located above the 20kK ZAMS extinction track with $\RV=3.1$. The arrow marks the  extinction direction for $\EBV = 1.0$~mag and the labels indicate the location of the RC, RGB, and AGB stars, as smeared by extinction, (right) Same CAMD after applying \textit{Gaia} + 2MASS filters to eliminate non-OB stars.}
  \label{method}
\end{figure}

Specifically regarding stellar groups with massive stars, the Villafranca project \citep{Maizetal20b,Maizetal22a,Maizetal25} is obtaining the membership of the most relevant targets with OB stars and properly deriving their distances using the calibration described in subsection 1.1. One of its most significant results is the discovery that the Bermuda cluster, the origin of the ionising stars of the North America and Pelican nebulae, is dissolving not because of the loss of its gas but because of three different dynamical events that ejected its most massive stars, leaving behind a expanding orphan cluster \citep{Maizetal22b}.

\section{Sample and spatial distribution}

$\,\!$ Up to DR3, the most important contribution of \textit{Gaia} to massive-star research has been the calculation of accurate and precise distances to stars in the solar neighborhood. This in turn has led to studies of their spatial distribution and of their sample, as it is now possible to build relatively accurate color-absolute magnitude diagrams (CAMDs) and from there to identify new massive stars to some degree. At the same time, an increased confusion in terminology has appeared. Traditionally, most\footnote{To these one has to add (most) A-M supergiants, (most) WR stars, and a zoo of other minor classes.} massive stars were identified as OB stars, meaning O-B2 dwarfs, O-B5 giants, and O-B8 supergiants, leaving the much more numerous mid- and late-B dwarfs out, as they are for the most part of intermediate mass \citep{Maizetal26}. However, \textit{Gaia} has prompted a number of searches for stars in the upper left part of the CAMD  \citep{Chenetal19b,Liuetal19b,Liuetal24,Poggetal21,Xuetal21,Zarietal21,LiBinn22,Xianetal22,Khaletal24,Quinetal25a} under names such as ``Upper Main Sequence stars'', ``OBA stars'', ``B-type candidates'', and even as ``OB stars'' when most of their sample is actually made out of late B dwarfs. The reader is referred to ALS~III \citep{Pantetal25b}, where we have compared those studies, indicating which ones are aware of that distinction and/or try to eliminate non-massive stars from the sample (with spectroscopy or otherwise) and which ones are/do not.

\begin{wrapfigure}{l}[12mm]{0.6\textwidth}
  \vspace{-5mm}
  \includegraphics[width=0.6\textwidth]{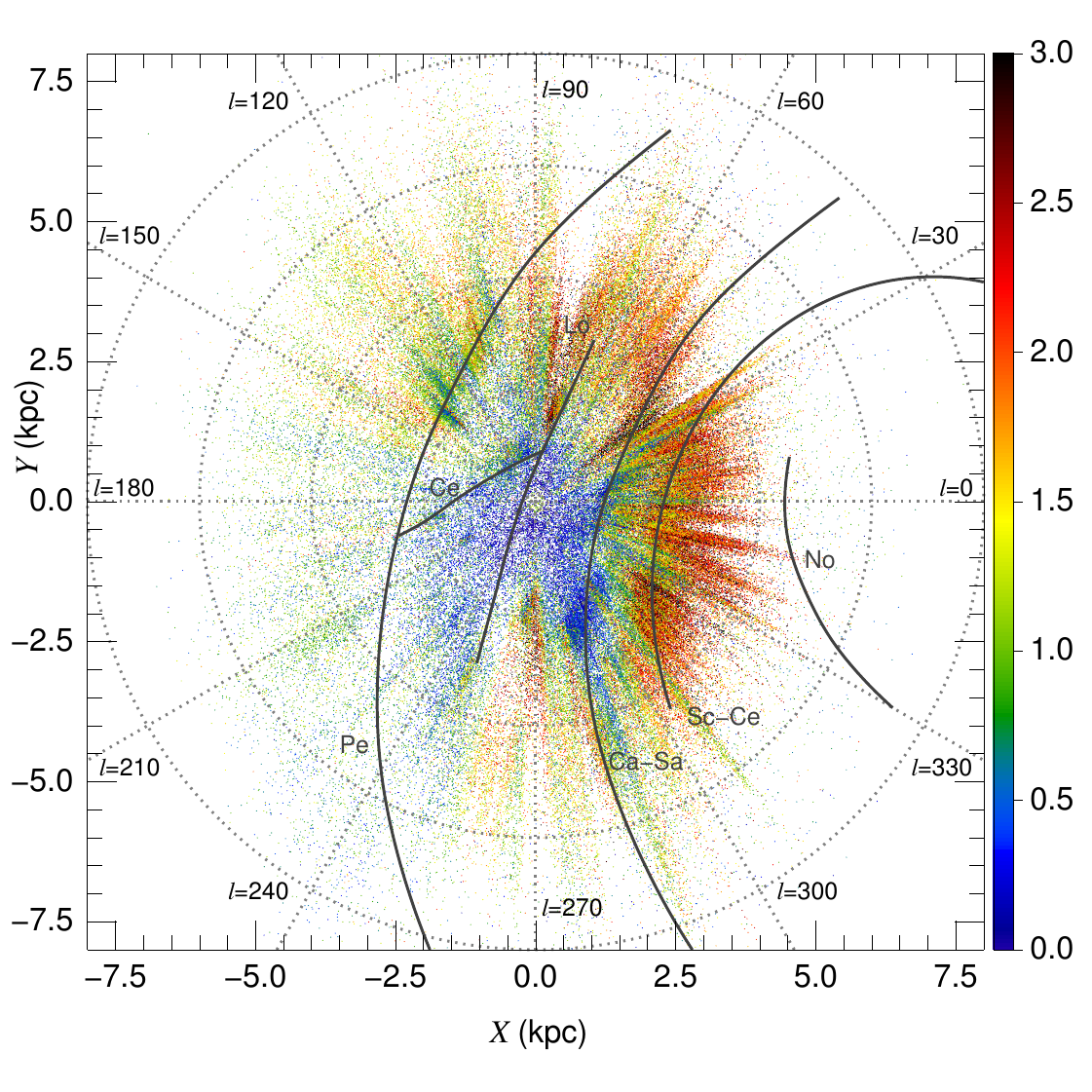}
  \vspace{-5mm}
  \caption{Spatial distribution of the sample in the right panel of Fig.~\ref{method} colour-coded by \GBPmGRP\ as a proxy for extinction.}
  \label{MWOB_xy}
\end{wrapfigure}

Most of the studies mentioned above easily detect the Carina-Sagittarius, Local, and Perseus spiral arms (the latter, only clearly seen for $l < 140^{\rm o}$, \citealt{NeguMarc08}). The \citet{Poggetal21} and \citet{Zarietal21} samples, though dominated by intermediate-mass  stars, also detect the Scutum-Centaurus spiral arm. \citet{Pantetal21} discovered the Cepheus arm as an interarm structure elevated over the Galactic plane. \citet{Kuhnetal21} detected a similar but shorter high-pitch angle structure in the Carina-Sagittarius arm.

ALS~III lists 15\,648 high-probability Galactic massive stars. Other studies that rely as ours on filtering with spectroscopy or other methods \citep{Chenetal19b,Xuetal21} reach similar or lower numbers. Others that are more lax on their selection criteria can reach values more than an order of magnitude higher\footnote{Some are explicit about their sample being dominated by non-massive stars, others are not.} (see Table~I in ALS~III). The question is that, given that the Galactic massive-star population is expected to be about two orders of magnitude higher than that, how can we use \textit{Gaia} to significantly increase the known sample? One answer (Ma\'{\i}z Apell\'aniz in prep.) is to combine \textit{Gaia} photometry and astrometry with 2MASS photometry to filter out the low+intermediate mass evolved population that dominates the upper right part of the Galactic CAMD in the form of red clump (RC), red giant branch (RGB), and asymptotic giant branch (AGB) stars. The effect of such a filter is seen by comparing the two panels in Fig.~\ref{method}: after applying the filters we are left with the expected distribution of a population of OB stars first extinguished and then limited in magnitude as we move towards redder colours. Even if the contamination from intermediate-mass stars is of the order of $50\%$, the filtering process yields $\sim 10^5$ stars. Figure~\ref{MWOB_xy} plots the filtered population projected onto the Galactic Plane, colour-coded by \GBPmGRP, where the increase in extinction at the location of the Scutum-Centurus arm is easily visible. With the better quality of DR4 data, the number of stars filtered by this method is expected to be several times higher, thus increasing the number of known Galactic massive stars by an order of magnitude from the current number.

\section{Summary}

$\,\!$ I have reviewed the previous and future contributions of \textit{Gaia} to the study of massive stars: parameter determination, binaries, photometric variability, stellar groups, and sample and spatial distribution. Of the previous ones, likely the most important one has been the ability to derive for the first time precise and accurate distances to massive stars within several kpc of the Sun, thus eliminating many uncertainties in their properties. Of the ones expected for DR4, two stand out: the likely discovery of a significant number of NS and BH companions and the increase of the sample of known Galactic massive stars by an order of magnitude.

\begin{adjmulticols}{2}{-5mm}{-8mm}
\bibliographystyle{aa}
\bibliography{general} 
\end{adjmulticols}

\end{document}